\documentclass[11pt]{article}

\makeatletter\renewcommand{\section}{\@startsection
{section}{1}{\z@}{-3.5ex plus -1ex minus
    -.2ex}{2.3ex plus .2ex}{\bf }}

\makeatletter\renewcommand{\subsection}{\@startsection{subsection}{2}{\z@}{-3.25ex
plus -1ex minus
   -.2ex}{1.5ex plus .2ex}{\it }}
\makeatletter\renewcommand{\subsubsection}{\@startsection{subsubsection}{3}{-2.45ex}{-3.25ex
plus -1ex minus -.2ex}{1.5ex plus .2ex}{\it }}
\renewcommand{\thesection}{\arabic{section}.}
\renewcommand{\thesubsection}{\arabic{section}.\arabic{subsection}.}

\renewcommand{\theequation}{\thesection\arabic{equation}}
\makeatletter \@addtoreset{equation}{section}

\usepackage{graphicx}
\usepackage{epsfig}

\newcommand{\be}{\begin{equation}}
\newcommand{\ee}{\end{equation}}
\newcommand{\bea}{\begin{array}}
\newcommand{\ea}{\end{array}}
\newcommand{\beqa}{\begin{eqnarray}}
\newcommand{\eeqa}{\end{eqnarray}}
\newcommand{\nn}{\nonumber}

\renewenvironment{thebibliography}[1]
     {\baselineskip=16pt plus 2pt minus 1pt
      \section*{\large\refname
        \@mkboth{\MakeUppercase\refname}{\MakeUppercase\refname}}%
     \list{\@biblabel{\@arabic\c@enumiv}}%
           {\settowidth\labelwidth{\@biblabel{#1}}%
            \leftmargin\labelwidth
            \advance\leftmargin\labelsep
            \@openbib@code
            \usecounter{enumiv}%
            \let\p@enumiv\@empty
            \renewcommand\theenumiv{\@arabic\c@enumiv}}%
      \sloppy
      \clubpenalty4000
      \@clubpenalty \clubpenalty
      \widowpenalty4000%
      \sfcode`\.\@m}

\let\fn\footnote
\renewcommand{\footnote}[1]{\linespread{1.1}\fn{#1}\linespread{1.29}}

\usepackage{amsfonts}
\usepackage{amssymb}
\usepackage{mathrsfs}
\usepackage{amsmath,amssymb}
\usepackage{bbm}
\usepackage{bm}

\newcommand{\appendices}{\section*{Appendix}\setcounter{subsection}{0} \setcounter{equation}{0}
\renewcommand{\thesubsection}{\Alph{subsection}.} 
\renewcommand{\theequation}{\thesubsection\arabic{equation}}
}

\def\tyng(#1){\hbox{\tiny$\yng(#1)$}}

\begin{document}

\begin{titlepage}
\begin{flushright}
ITP--UH--18/07\\
\end{flushright}
\vskip 2.0cm

\begin{center}

\centerline{{\Large \bf Quantum Aspects of the Noncommutative Sine-Gordon Model}} 

\vskip 5em

\centerline{\large \bf Se\c{c}kin~K\"{u}rk\c{c}\"{u}o\v{g}lu  and Olaf Lechtenfeld}

\vskip 2em

{\small
\centerline{\sl Institut f\"ur Theoretische Physik, Leibniz Universit\"at Hannover}
\centerline{\sl Appelstra\ss{}e 2, D-30167 Hannover, Germany} 
     
\vskip 1em

{\sl  e-mails:}  \hskip 2mm {\sl seckin.kurkcuoglu@itp.uni-hannover.de \,,  lechtenf@itp.uni-hannover.de} 
}
\end{center}

\vskip 3cm

\begin{quote}
\begin{center}
{\bf Abstract}
\end{center}
\vskip 5pt
In this paper, we first use semi-classical methods to study quantum field theoretical aspects of the integrable
noncommutative sine-Gordon model proposed in [hep-th/0406065]. In
particular, we examine the fluctuations at quadratic order around the
static kink solution using the background field method. We derive
equations of motion for the fluctuations and argue that at $O
(\theta^2)$ the spectrum of fluctuations remains essentially the same
as that of the corresponding commutative theory. We compute the one-loop two-point 
functions of the sine-Gordon field and the additional scalar 
field present in the model and exhibit logarithmic divergences, which lead to UV/IR mixing.
We briefly discuss the one-loop renormalization in Euclidean signature and comment on the obstacles in determining the noncommutativity corrections to the quantum mass of the kink.
\end{quote}

\vskip 3cm
\begin{quote}
\end{quote}

\end{titlepage}

\setcounter{footnote}{0}

\newpage

\section{Introduction}

Several aspects of classical and quantum field theories on noncommutative deformations of spacetime have been under investigation for some time now. Among them, field theories defined on Groenewold-Moyal type deformations of $3+1$ and $2+1$ dimensional spacetime hold a considerably large part of the literature (see for example \cite{Douglas} for a review), whereas theories in $1+1$ dimensions have not been considered extensively until very recently. In \cite{NCSG0, NCSG01} a noncommutative deformation of the sine-Gordon model was constructed, however it lacked some of the desirable features of a $1+1$ dimensional field theory even at the classical level. 

In \cite{NCSG1} a novel noncommutative deformation of the sine-Gordon model has been proposed. This model is obtained through a dimensional reduction of a certain integrable time-space noncommutative sigma model in $2+1$ dimensions, which was previously constructed in \cite{Lechtenfeld:2001gf}. In \cite{NCSG1} it was demonstrated that this particular deformation of the sine-Gordon model possesses many attractive features at the classical level, as would be expected from a theory in $1+1$ dimensions. Firstly, its classical integrability is guaranteed as it is obtained by dimensional reduction from the linear system of the noncommutative integrable sigma model. Although this dimensional reduction takes place initially at the level of equations of motion, it also works at the level of the action, leading directly to the sought for action in $1+1$ dimensions.
It was also shown in \cite{NCSG1} that solitonic solutions of the model exist, and the presence of the linear system made it possible to use the well-known technique of ``dressing'' to find these solutions in a systematic manner. Finally, a direct evaluation of the tree-level amplitudes performed in \cite{NCSG1} revealed that the theory has a causal $S$-matrix and that no particle production occurs.

All these features make this model quite an attractive testing ground for launching further investigations on noncommutative deformations of $1+1$-dimensional field theories. In particular, it is desirable to find some indications on the behaviour of this model as a quantum field theory. With this state of mind, we first investigate the fluctuation spectrum in the background of the one-kink solution by applying elementary semi-classical methods. We find that at $O(\theta)$ in perturbation theory the spectrum of quadratic fluctuations remains the same as that for the commutative sine-Gordon theory.
We also argue that this spectrum remains essentially the same at order $O (\theta^2)$ as well. 

Next, we turn our attention to the two-point functions of the sine-Gordon field and the additional scalar field, in the vacuum sector at one-loop order, and exhibit that they both have logarithmic divergences. The amplitude for the sine-Gordon field also contains a piece which leads to UV/IR mixing. Interestingly, both two-point functions receive contribution from loop integrals which arise only from noncommutativity but do not lead to UV/IR mixing. Finally, we discuss the renormalization of the model for the Euclidean signature and comment on the obstacles in determining the noncommutativity corrections to the quantum mass of the kink. 
 
\section{Basics}

In this section, we collect some elementary definitions to set the notation and conventions used throughout the text. We work in the $1+1$-dimensional Groenewold-Moyal spacetime ${\cal A}_\theta({\mathbb R}^{1+1})$, generated by the coordinates $t$ and $y$ with the commutation relations
\be
\lbrack t \,, y \rbrack_\star : =  t \star y - y \star t = i \theta \,.
\ee
The star product is defined by
\be
(\alpha \star \beta) (t, y) = \alpha \, e^{\frac{i}{2} \theta (\overleftarrow{\partial}_t \overrightarrow{\partial}_y - \overleftarrow{\partial}_y \overrightarrow{\partial}_t)} \, \beta \,, \quad \alpha, \beta \in {\cal A}_\theta({\mathbb R}^{1+1}) \,.
\ee

In order to avoid cluttered notation, we suppress the $\star$ notation for the star products in all the formulae from now on. It is also understood that functions such as $e^{f(t, y)}$ stand for $e^{f(t,y)}_\star:= 1 + f + \frac{1}{2} f \star f + \cdots$. Throughout the paper, it will always be clear from the context whether the star product or the pointwise product is involved in a formula.

Let $g_+$ and $g_-$ be two elements of ${\cal A}_\theta({\mathbb R}^{1+1})$, which are valued in $U(1)_\star$. Then the noncommutative sine-Gordon model of reference \cite{NCSG1} can be defined by the action functional
\be
S[g_+, g_-] = S_{WZW}[g_+] +  S_{WZW}[g_-] + \alpha^2 \int dt dy \, (g_+^\dagger g_- + g_-^\dagger g_+ -2) \,,
\label{eq:NCSGaction}
\ee
where
\be
S_{WZW}[f] = - \frac{1}{2} \int dt dy \, \partial_\mu f^{-1} \partial^\mu f - \frac{1}{3} \int dt dy \int_0^1 d \lambda \varepsilon^{\mu\nu\sigma}\, {\hat f}^{-1} \partial_\mu {\hat f} {\hat f}^{-1} \partial_\nu {\hat f} {\hat f}^{-1} \partial_\sigma {\hat f} \,.
\label{eq:WZW}
\ee
The Wess-Zumino (WZ) term in (\ref{eq:WZW}) contains a path on the interval $\lbrack 0 \,, 1 \rbrack$, parametrized by a coordinate $\lambda$, which commutes with both $t$ and $y$. ${\hat f} (t, y, \lambda)$ is an extension of $f(t, y)$ on this interval, interpolating between
\be
{\hat f} (t, y , 0) = \mbox{constant} \,, \quad {\hat f} (t, y, 1) = f(t, y) \,.
\ee

It is possible to parametrize $g_\pm$ in terms of scalar fields $\phi_\pm(t, y)$ as 
\be
g_+ = e^{-i \phi_+} \,, \quad g_- = e^{ i \phi_-} \,.
\ee

Taking $\theta \rightarrow 0$ and using the field redefinitions $\varphi := \phi_+ + \phi_-$ and $\rho := \phi_+ - \phi_-$, the action $S[g_+, g_-]$ leads to the usual (commutative) sine-Gordon action in the field $\varphi$ plus a free scalar field action for the field $\rho$.

For further details on this model and its derivation from a certain noncommutative sigma model in $2+1$ dimensions we refer the reader to the original articles \cite{NCSG1, Chu:2005tv}.

\section{Fluctuations around a classical background}

\subsection{Stability equations}

We split the fields by setting
\be
g_+ = g_{0 +} e^{-i \pi_+} \,, \quad \quad g_- = e^{i \pi_-} g^{-1}_{0 -} \,,
\ee
where the set $\lbrace g_{0 +} \,, g_{0 -} \rbrace$ is any background satisfying the classical equations of motion that follow from $S[g_+, g_-]$, and $\pi_+ \,, \pi_-$ are the fluctuations in this background. In the following subsections we will examine the vacuum and the one-kink solutions as backgrounds, which are both static. In any static background, one has
\be
g_{0 +} = g_{0 -} = g_0  \quad \Longleftrightarrow \quad \phi_+ = \phi_- = : \phi_0
\ee
and from now on we will restrict ourselves to such backgrounds.
 
We expand the action $S[g_+, g_-]$ to quadratic order in the fluctuations $\pi_\pm$. A long but straightforward calculation gives
\begin{multline}
S[g_+, g_-] =  \int dt dy \, \Big \lbrack - \frac{1}{2} \partial_\mu g_0^{-1} \partial^\mu g_0 - \frac{1}{2} (\partial_\mu \pi_+)^2 
- \frac{1}{2} (\partial_\mu \pi_-)^2  \\
- \Big (\frac{1}{2} \eta^{\mu \nu} + \varepsilon^{\mu \nu} \Big) g_0^{-1} \partial_\mu g_0 
\Big ( \lbrack \partial_\nu \pi_+ \,, \pi_+ \rbrack + \lbrack \partial_\nu \pi_- \,, \pi_- \rbrack \Big ) \Big \rbrack \\
+ \alpha^2 \int dt dy  \,  \, \Big \lbrack g_0^{-2} + g_0^2 
- \frac{1}{2} (\pi_+^2 + \pi_-^2) (g_0^{-2} + g_0^2) - \pi_+ g_0 \pi_- g_0 -  \pi_+ g_0^{-2} \pi_ - g_0^{-2} \Big \rbrack + O(\pi^3)  \,,
\end{multline}
up to cubic and higher order terms in $\pi_\pm$. This leads to the following equations of motion for $\pi_\pm$:
\begin{multline}
-\partial_\mu \partial^\mu \pi_\pm + (\eta^{\mu \nu} - 2 \varepsilon^{\mu \nu}) \Big ( \lbrack \partial_\mu \pi_\pm \,, 
g_0^{-1} \partial_\nu g_0 \rbrack + \frac{1}{2} \lbrack \pi_\pm \,, \partial_\mu (g_0^{-1} \partial_\nu g_0) \rbrack \Big ) \\
- \frac{\alpha^2}{2} \lbrace \pi_\pm \,, g_0^{-2} + g_0^2 \rbrace - \alpha^2 ( g_0 \pi_\mp g_0 + g_0^{-1} \pi_\mp g_0^{-1}) + O(\pi^2) = 0 \,.
\label{eq:coupledeq}
\end{multline}
In (\ref{eq:coupledeq}) square and curly brackets denote respectively the commutators and anticommutators with respect to the star product. In the following we will work at order linear in $\pi_\pm$.

The equations in (\ref{eq:coupledeq}) decouple if we redefine the fluctuating fields as  
\be
\eta := \frac{1}{2} (\pi_+ + \pi_-) \,, \quad \quad \xi := \frac{1}{2} (\pi_+ - \pi_-) \,.
\ee
Thus we have
\begin{multline}
-\partial_\mu \partial^\mu \eta + (\eta^{\mu \nu} - 2 \varepsilon^{\mu \nu}) \Big ( \lbrack \partial_\mu \eta \,, 
g_0^{-1} \partial_\nu g_0 \rbrack + \frac{1}{2} \lbrack \eta \,, \partial_\mu (g_0^{-1} \partial_\nu g_0) \rbrack \Big ) \\
- \frac{\alpha^2}{2} \lbrace \eta \,, g_0^{-2} + g_0^2 \rbrace - \alpha^2 ( g_0 \eta g_0 + g_0^{-1} \eta g_0^{-1}) = 0 \,,
\label{eq:eta}
\end{multline}
\begin{multline}
-\partial_\mu \partial^\mu  \xi + (\eta^{\mu \nu} - 2 \varepsilon^{\mu \nu}) \Big ( \lbrack \partial_\mu  \xi \,, 
g_0^{-1} \partial_\nu g_0 \rbrack + \frac{1}{2} \lbrack  \xi \,, \partial_\mu (g_0^{-1} \partial_\nu g_0) \rbrack \Big ) \\
- \frac{\alpha^2}{2} \lbrace  \xi \,, g_0^{-2} + g_0^2 \rbrace + \alpha^2 ( g_0  \xi g_0 + g_0^{-1}  \xi g_0^{-1}) = 0 \,.
\label{eq:xi}
\end{multline}
Let us now examine the consequences of (\ref{eq:eta}) and (\ref{eq:xi}) in the vacuum and one-kink sector.

\subsection{Fluctuations in the vacuum sector}

In this case we have
\be
g_0 = e^{-\frac{i}{2} \varphi_0} = 1  \,, \quad \varphi_0 = 0 \,, \quad \rho_0 = 0 \,,
\ee
hence (\ref{eq:eta}) and (\ref{eq:xi}) simplify to
\be
-\partial_\mu \partial^\mu \eta - 4 \alpha^2 \eta = 0 \,, \quad -\partial_\mu \partial^\mu \xi = 0 \,.
\ee
Thus, in the vacuum background, the fluctuations $\eta$ and $\xi$ are plane waves
\be
\eta(t, y) = e^{\pm i k y + i \omega t} \,, \quad  \xi(t, y) = e^{\pm i r y + i \nu t} \,,
\ee
with the dispersion relations $\omega^2 = k^2 + 4 \alpha^2$ and $\nu^2 = r^2$. These results are in complete agreement with 
the fluctuation spectrum in the vacuum sector of the usual sine-Gordon theory. The presence of the $\xi$-fluctuations does not effect this conclusion as they are decoupled from $\eta$ in this background.
 
\subsection{Fluctuations in the kink sector}

Let us now examine the static one-kink solution for the $g_0$ background. In this case we have
\be
g_0 = e^{-\frac{i}{2} \varphi_0} \,, \quad \varphi_0 = 4 \arctan e^{-2 \alpha y} \,,  \quad \rho_0 = 0 \,.
\label{eq:kink}
\ee
We observe that (\ref{eq:eta}) and (\ref{eq:xi}) are complicated equations in which infinitely many derivatives in time and space appear due to the star product. It does not seem possible to solve these equations analytically. In order to extract some physical information from these equations, let us assume that the noncommutativity is rather small and allows us to expand the star product in powers of $\theta$\footnote{In \cite{Wimmer} the fluctuation spectrum of noncommutative Yang-Mills instantons are studied without performing a $\theta$ expansion. It would be worthwhile to investigate the adaptability of the methods of \cite{Wimmer} to the current model.}. 

Expanding (\ref{eq:eta}) and (\ref{eq:xi}) to second order in $\theta$, we find (disregarding $0(\theta^3)$ terms)
\begin{multline}
-\partial_\mu \partial_\mu \eta - 4 \alpha^2 \cos  \varphi_0 + \frac{1}{2} \theta \partial_y^2  \varphi_0 \partial_t \partial_y \eta 
+ \frac{\theta}{4} \partial_y^3  \varphi_0  \partial_t \eta + \theta \partial_y^2  \varphi_0 \partial_t^2 \eta \\
-\frac{1}{2} \alpha^2 \theta^2 \Big (\partial_y^2  \varphi_0 \sin  \varphi_0 + (\partial_y  \varphi_0)^2 \cos  \varphi_0 \Big ) \partial_t^2 \eta = 0 \,,
\label{eq:eta2}
\end{multline}
\be
-\partial_\mu \partial_\mu \xi + \frac{1}{2} \theta \partial_y^2  \varphi_0 \partial_t \partial_y \xi
+ \frac{\theta}{4} \partial_y^3  \varphi_0  \partial_t \xi + \theta \partial_y^2  \varphi_0 \partial_t^2 \xi = 0 \,,
\label{eq:xi2}
\ee
We now assume the following mode expansion for the fluctuations,
\be
\eta(t, y) = \sum_n e^{i \omega_n t} \psi_n(y) \,, \quad \xi(t, y) = \sum_n e^{i \nu_n t} \chi_n(y) \,.
\ee
Substituting these into (\ref{eq:eta2}) and (\ref{eq:xi2}) and projecting to an eigenmode labelled by $n$ we find
\beqa
\partial_y^2 \psi_n(y) + A \partial_y \psi_n(y) + B \psi_n(y) = 0 \,, \nn \\
\partial_y^2 \chi_n(y) + C \partial_y \chi_n(y) + D \chi_n(y) = 0 \,,
\label{eq:partialdiff1}
\eeqa
where $A, B, C, D$ are given by
\beqa
A &=& \frac{i}{2} \omega_n \theta \partial_y^2 \varphi_0 \,, \nn \\
B &=& \Big (1 - \theta \partial_y^2 \varphi_0 + \frac{1}{2} \alpha^2 \theta^2 (\partial_y^2 \varphi_0 \sin \varphi_0 + (\partial_y \varphi_0)^2 \cos \varphi_0 ) \Big ) \omega_n^2 + \frac{i}{4} \theta \omega_n \partial_y^3 \varphi_0 - 4 \alpha^2 \cos \varphi_0 \,, \nn \\ 
C &=& \frac{i}{2} \nu_n \theta \partial_y^2 \varphi_0 \,, \nn \\
D &=& (1 - \theta \partial_y^2 \varphi_0 ) \nu_n^2 + \frac{i}{4} \theta \nu_n \partial_y^3 \varphi_0 \,, 
\eeqa
Making the substitutions
\be
\psi_n := e^{- \frac{i}{4} \omega_n \theta \partial_y \varphi_0} {\tilde \psi}_n \,, \quad 
\chi_n := e^{- \frac{i}{4} \nu_n \theta \partial_y \varphi_0} {\tilde \chi}_n \,
\ee
is sufficient to eliminate the terms which are first order in the $y$-derivatives in (\ref{eq:partialdiff1}) and cast them into
\beqa
\partial_y^2 {\tilde \psi}_n(y) + \Big ( B - \frac{1}{4}A^2 - \frac{1}{2} \partial_y A \Big ) {\tilde \psi}_n(y) = 0 \,, \nn \\
\partial_y^2 {\tilde \chi}_n(y) + \Big ( D - \frac{1}{4}C^2 - \frac{1}{2} \partial_y C \Big ) {\tilde \chi}_n(y) = 0 \,.
\label{eq:partialdiff2}
\eeqa
Using (\ref{eq:kink}) and defining $z := 2 \alpha y$ we can write (\ref{eq:partialdiff2}) as
\begin{multline}
-\partial_z^2 {\tilde \psi}_n(z) + (2 \tanh^2 z - 1) {\tilde \psi}_n(z) \\ 
- \Big ( 2 \theta \omega_n^2 \frac{\sinh z}{\cosh^2 z}  
+ \omega_n^2 \alpha^2 \theta^2 \big ( \frac{2}{\cosh^4 z} - \frac{\sinh^2 z}{\cosh^4 z} \big )\Big ) {\tilde \psi}_n(z) = \frac{\omega_n^2}{4 \alpha^2} {\tilde \psi}_n(z) \,,
\label{eq:potential1}
\end{multline}
\be
-\partial_z^2 {\tilde \chi}_n(z) - \Big ( 2 \theta \nu_n^2 \alpha^2 \frac{\sinh z}{\cosh^2 z} 
-\omega_n^2 \alpha^2 \theta^2 \frac{\sinh^2 z}{\cosh^4 z} \Big ) {\tilde \chi}_n(z) = \frac{\nu_n^2}{4 \alpha^2} {\tilde \chi}_n(z) \,.
\label{eq:potential2}
\ee

Invoking the standard semi-classical reasoning (see for example \cite{Rajaraman}, \cite{Rebhan}), we can write the energy spectrum in the kink sector up to order $O(\alpha^2)$ as
\be
E_{kink-sector} = 16 \alpha + \frac{1}{2} \sum_n (\omega_n + \nu_n) + O(\alpha^2) \,.
\label{eq:kink1}
\ee
Note that in this expression the frequencies $\nu_n$ associated to the field $\rho$ also appear, as the kink sector is specified by the configuration (\ref{eq:kink}).

Hence, we now have the task of determining the eigenvalues $\omega_n$ and $\nu_n$. (\ref{eq:potential1}) and (\ref{eq:potential2}) are one dimensional Schr\"{o}dinger-type equations with complicated ``potentials''. However, exact solutions for these equations are known when $\theta = 0$. Thus, we may treat the $\theta$ dependent terms as perturbations and $\theta$ as a perturbation parameter. We now investigate different cases in some detail. \\[0.5em]
{\it $\theta \rightarrow 0 $ limit}: \\
In this case (\ref{eq:potential1}) and (\ref{eq:potential2}) reduce to
\beqa
-\partial_z^2 \, {}_0\psi_n(z) + (2 \tanh^2 z - 1) \, {}_0\psi_n(z) &=& \frac{{}_0\omega_n^2}{4 \alpha^2} \, {}_0\psi_n(z) \,,
\label{eq:potc1}
\\
-\partial_z^2 \, {}_0\chi_n(z) &=& \frac{{}_0\nu_n^2}{4 \alpha^2} \, {}_0\chi_n(z) \,.
\label{eq:potcom2}
\eeqa
where the left subscript in ${}_0\psi_n(z) \,, {}_0\omega_n^2$ etc. are put to indicate that they are the corresponding objects at $\theta = 0$.

We recognize (\ref{eq:potc1}) as the equation of quadratic fluctuations of the usual (commutative) sine-Gordon theory. 
It belongs to the class of Schr\"{o}dinger-type equations with reflectionless potentials \cite{Morse} and has the discrete zero-mode solution 
\be
{}_0\psi_0(z) = \partial_z \varphi_0 = - \frac{2}{\cosh z}  \,, \quad \omega_0 = 0 \,,
\ee
followed by a continuum of states 
\be
{}_0\psi_q(z) = e^{i q z} (\tanh z - i q) \,, \quad  {}_0\omega_q^2 = 4 \alpha^2 (q^2 +1) \,, \quad q \geq 0  \,.
\ee
Usually, ${}_0\psi_q(z)$ are normalized by imposing periodic boundary conditions ${}_0\psi_q(z + \frac{L}{2}) = {}_0\psi_q(z - \frac{L}{2})$ in a box of length $L$. Fluctuations in the vacuum sector can be normalized likewise. These require
\be
q_n 2 \alpha L \delta(q_n) = 2 \pi n = k_n L \,,
\label{eq:periodic}
\ee
where $ \delta(q_n)$ is the phase shift defined below. The normalized states are
\be
{}_0\psi_{q_n}(z) = N e^{i q_n z} (\tanh z - i q_n) \,, \quad N = L + L q^2 - 2 \tanh \frac{L}{2} \,,
\label{eq:normalization} 
\ee
where $N$ is the normalization factor. ${}_0\psi_{q_n}(z)$ has the asymptotic behaviour
\be
{}_0\psi_{q_n}(z) \longrightarrow e^{i q_n z} e^{\pm \frac{1}{2} \delta(q_n)} \,,
\ee
and
\be   
\delta(q_n) = \pi {\rm sgn}(q_n) - 2 \arctan q_n \,
\ee
is the associated phase shift. 

The equation (\ref{eq:potcom2}) is trivially solved by 
\be
{}_0\chi_n(z) = e^{\pm i \frac{\nu_n}{2 \alpha} z } \,.
\ee
Thus, the fluctuations ${}_0\xi$ are plane waves as expected since, at $\theta = 0$, ${}_0\xi$ represent the fluctuations   
of the scalar field $\rho$, which is free in this limit. \\[0.5em]
{\it Zero-mode:} \\
We observe that 
\be
\psi_0(z) = \partial_z \varphi_0 = - \frac{2}{\cosh z} \,
\ee
is a solution of (\ref{eq:eta}) with zero frequency. This can be verified easily by direct substitution of $\partial_z \varphi_0$ in (\ref{eq:eta}). Thus, the only discrete mode of the commutative theory is unaffected by the presence of noncommutativity. In fact, this conclusion can also reached by noting that the kink solution and the associated zero mode are both independent of the time coordinate, thus all star products collapse to pointwise products.\\[0.5em]
{\it Perturbation theory:} \\
Let us treat $\theta$ in (\ref{eq:potential1}) and (\ref{eq:potential2}) as the perturbation parameter. For the consistency of this assumption we further require that $\omega_n$ and $\nu_n$ dependence of the terms at order $\theta$ and $\theta^2$ in (\ref{eq:potential1}) and (\ref{eq:potential2}) are approximated by the commutative spectrum ${}_0\omega_n$ and ${}_0\nu_n$.

In order to apply standard perturbation theory, we put the entire system in a box of length $L$, so that both $\omega_n$  and $\nu_n$ have discrete spectra. When $L \rightarrow \infty$, the continuum structure will be recovered. Let us focus on the spectrum of $\omega_n^2$. The potential is read off from (\ref{eq:potential1}) to be
\begin{multline}
V(z) =  (2 \tanh^2 z - 1) - 2 \theta \omega_n^2 \frac{\sinh z}{\cosh^2 z} + \omega_n^2 \alpha^2 \theta^2 \Big ( \frac{2}{\cosh^4 z} - \frac{\sinh^2 z}{\cosh^4 z} \Big ) \\[1em]
: = V_0(z) + \theta V_1(z) + \theta^2 V_2(z) \,
\end{multline}
and depends on the modes $\omega_n$ themselves. Symbolically we can express the corrections to the spectrum of $\omega_n^2$ as
\be
\omega_n^2 - {}_0\omega_n^2 =: \Delta_n(V_1) + \Delta_n(V_2)  \,, 
\ee
where
\beqa
\Delta_n(V_1) &=& \theta \Delta_n^{(1)}(V_1) + \theta^2 \Delta_n^{(2)}(V_1) + \cdots \,, \nn \\
\Delta_n(V_2) &=& \theta^2 \Delta_n^{(1)}(V_2) + \theta^4 \Delta_n^{(2)}(V_2) + \cdots \,,
\eeqa
and the superscripts indicate the order of perturbation theory.
Applying the perturbation theory at first-order in $\theta$, we immediately observe that corrections at this order vanish: $\Delta_n^{(1)}(V_1) = 0$, since $V_1(z)$ is odd under parity $z \rightarrow -z$.

Let us move on to discuss the corrections at order $\theta^2$. In this case, it is sufficient to treat the terms of order $\theta^2$ in first-order perturbation theory, while it is necessary to apply second-order perturbation theory to terms of order $\theta$. 
Applying first-order perturbation theory to $V_2(z)$ gives:
\be
\Delta_n^{(1)}(V_2) = {}_0\omega_n^2 \alpha^2 |N|^2 \int d z \, |\tanh z - iq_n|^2  \Big ( \frac{2}{\cosh^4 z} - \frac{\sinh^2 z}{\cosh^4 z} \Big ) \,,
\label{eq:expectation1}
\ee
where $N$ is the normalization factor given in (\ref{eq:normalization}). The integral in (\ref{eq:expectation1}) can be computed exactly. To leading order in $L$ we find 
\beqa
\Delta_n^{(1)}(V_2) &\approx & {}_0\omega_n^2 \alpha \left (\frac{1}{15} + q_n^2 \right) \frac{1}{L} \,, \nn \\
&\approx & {}_0\omega_n^2 \alpha \left (\frac{1}{15} + \frac{k_n^2}{4 \alpha^2}  \right) \frac{1}{L} + O(\frac{1}{L^2}) 
\eeqa
since it follows from (\ref{eq:periodic}) that $q_n^2 = \frac{1}{4 \alpha^2} (k_n^2 - \frac{2 \delta(k_n)}{L}) + O(\frac{1}{L^2})$. Thus $\Delta_n^{(1)}(V_2)$ vanishes in the limit $ L \rightarrow \infty$. 
\begin{figure}  
\centering   
\includegraphics[width=0.7\textwidth, height=0.3\textheight]{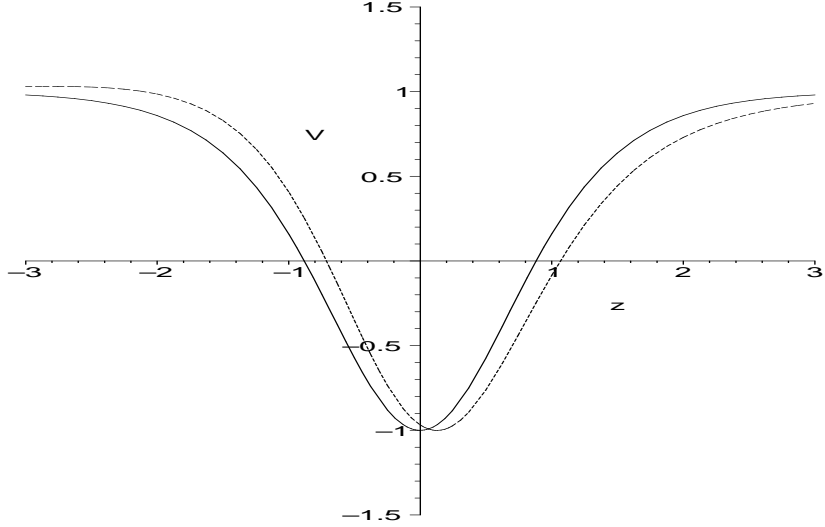}
\caption{``Potential'' $V(z)$ for (\ref{eq:potential1}). Solid line represents $V(z)$ at $\theta = 0$. Dashed line is at the values $\theta \alpha^2 = \frac{1}{16}$ and ${}_0\omega_q^2 = 4 \alpha^2$.}
\end{figure}

Second-order perturbation theory is required to determine $\Delta_n^{(2)}(V_1)$. However, its calculation becomes too complicated to extract an analytical answer even in the large-$L$ limit. Perhaps a numerical study could help to assess the strength of this term as $L \rightarrow \infty$. Nevertheless, we observe that the perturbing potentials $V_1(z)$ and $V_2(z)$ both fall off to zero exponentially fast as $z \rightarrow \pm \infty$, and $V(z)$ converges to one in both these limits (see, Fig. 1). These considerations suggest that the starting point $\omega_n^2 = 4 \alpha^2$ of the continuous spectrum remains unaltered, while the density of states are probably stirred up to a degree which is insensitive to the methods applied in this paper. Thus it seems rather unlikely that $\Delta_n^{(2)}(V_1)$ will substantially alter the spectrum of fluctuations. 

Similar statements are also valid for the perturbing potential in (\ref{eq:potential2}). In particular, corrections to first order in $\theta$ vanish, since this perturbation is also odd under parity. Thus the dispersion relation for the fluctuations $\xi$ remains the same as that of the vacuum sector $\nu^2_n = r_n^2$.  

After this analysis, it is now possible to perform the vacuum subtraction from $E_{kink}$ by writing
\beqa
E_{kink} - E_{vac}  &=& 16 \alpha + \frac{1}{2} \sum_n 2 \alpha (q_n^2 +1)^{\frac{1}{2}} + \sum_n r_n - \frac{1}{2} \sum_n (k^2_n + 4 \alpha^2)^{\frac{1}{2}} - \sum_n r_n + O(\alpha^2) \nn \\
&=& 16 \alpha - \frac{1}{4 \pi}  \int d k \sqrt{k^2 + 4 \alpha^2} \frac{d}{d k} \delta(k) + O(\alpha^2) + O(\theta^3) \,.
\label{eq:vacsubtract}
\eeqa
Up to order $O(\theta^3)$ this coincides with the usual expression for the sine-Gordon model \cite{Dashen, Rajaraman}, as we have argued that $q_n$ remains unaltered at order $\theta^2$. Finally, we note that in the corresponding commutative model, a mathematically precise treatment of the vacuum energy subtraction and alternative methods for regularization of the remaining divergences are presented in \cite{Rebhan}. 

\section{One-loop two-point functions}

In this section we compute the two-point functions for the sine-Gordon field and the additional scalar field in the model at the one-loop level in the vacuum sector. We observe that for this purpose it will suffice to know the action $S[g_+, g_-]$ to quadratic order in the fields $\phi_\pm$. Making this expansion and performing the field redefinitions
\be
\varphi : = \phi_+ + \phi_- \,, \quad \rho : = \phi_+ - \phi_- \,,
\ee
we find
\begin{eqnarray}
&& S[\varphi \,, \rho] = - \frac{1}{4} \partial_\mu \varphi \partial^\mu \varphi - \frac{1}{4}
\partial_\mu \rho \partial^\mu \rho - \frac{1}{4!} \frac{1}{2^3} \Big
( \lbrack \partial_\mu \varphi \,, \varphi \rbrack^2 + \lbrack
\partial_\mu \rho \,, \rho \rbrack^2 + 2 \lbrack \partial_\mu \varphi \,, \varphi \rbrack 
\, \lbrack \partial^\mu \rho \,, \rho \rbrack \Big ) \nn \\    
&& \quad \quad \quad \quad  \quad - \frac{i}{4} \varepsilon^{\mu \nu} ( 3 \partial_\mu \varphi
\partial_\nu \varphi \rho + \partial_\mu \rho \partial_\nu \rho \rho )
+ \alpha^2 \big ( - \varphi^2 + \frac{1}{12} \varphi^4 \big ) + O(\varphi^k \rho^{5-k}) \,.
\label{eq:quadraticaction}
\end{eqnarray}
From the commutative limit of (\ref{eq:NCSGaction}) or (\ref{eq:quadraticaction}) it is clear that $\varphi$ is the sine-Gordon field.
Feynman rules are extracted from this action, and they are listed in Appendix A. 

For $\varphi$, we find that one-loop two-point function 
\be
\langle \varphi(P) \varphi (P) \rangle : = I_\varphi(P^2)
\ee
in momentum space is given by the sum of the following integrals:
\be
I_1 =  \frac{2 \alpha^2 }{3 \, (2 \pi)^2}\int d^2 k \frac{2}{k^2 + 4 \alpha^2} \,, \quad  
I_2 =  \frac{\alpha^2}{3 \, (2 \pi)^2}\int d^2 k \frac{2}{k^2 + 4 \alpha^2} e^{- i \theta k \wedge P } \,,
\label{eq:I12}
\ee
\be
I_3 = \frac{3^2}{2^2 \, (2 \pi)^2} \int d^2 k \frac{(k \wedge P)^2 \sin^2 \left (\theta 
\frac{k \wedge P}{2} \right )}{(k^2 + 4 \alpha^2)((k-P)^2 + \mu^2)} \,,
\label{eq:I3}
\ee
\be
I_4 = \frac{-i}{2^3 \, (2 \pi)^2} \int d^2 k \frac{2}{k^2 + 4 \alpha^2} \Big (k^2 e^{- i \theta \frac{k \wedge P}{2}} - 
P^2 e^{i \theta \frac{k \wedge P}{2}} \Big ) \sin (\theta \frac{k \wedge P}{2}) \,.
\label{eq:I4}
\ee
In $I_3$ a small mass $\mu$ for the field $\rho$ has been introduced
to regularize the $IR$ divergence of this integral. 
We have also used the $\wedge$ symbol, which is defined by 
\be
a \wedge b := a_t b_y - a_y b_t \,.
\ee
The integrals can be performed by standard methods, and full results are given in Appendix B. Up to
leading order in $\theta$ and the momentum cut-off $\Lambda$, we find the following results, depending on the external momentum $P$ being zero or not. \\[0.2em] 
{\it For $P = 0$:} \\
$I_3$ and $I_4$ vanish while $I_1$ and $I_2$ add up to give
\be
I_\varphi(P^2=0) = \frac{-\alpha^2}{2 \pi} \log \frac{4 \alpha^2}{\Lambda^2} + \mbox{subleading terms (s.t.)} \,.
\ee
In this case, the result coincides with that of the usual sine-Gordon model at one loop. \\[0.2em] 
{\it For $P \neq 0$:} \\
\be
I_1 = \frac{-\alpha^2}{3 \pi} \log \frac{4 \alpha^2}{\Lambda^2} + \mbox{s.t.} \,, \quad 
I_2 = \frac{-\alpha^2}{6 \pi} \log \left [ \alpha^2 \theta^2 P^2 + \frac{4 \alpha^2}{\Lambda^2} \right ] + \mbox{s.t.} \,,
\ee
\begin{multline}
I_3 = \frac{3^2}{2^6 \pi} \Bigg [ - \frac{8}{\theta} + \int_0^1 d x \Big ( P^2 \log \frac{\theta^2 P^2 A}{4} - P^2 \log \frac{4 \alpha^2}{\Lambda^2} - P^2 \log \Big [1 + \frac{P^2}{4\alpha^2}x(1-x) \Big ] \\
- 2 \theta P^2 A  \log \Big [\frac{\theta  P}{2} \sqrt{A} \Big ]  \Big ) \Bigg ] + \mbox{s.t.} \,,
\label{eq:II3}
\end{multline}
\be
I_4 = \frac{1}{2^5 \pi} \left ( - \frac{2}{\theta^2 P^2} - P^2 \log
(\alpha^2 \theta^2 P^2 ) + P^2 \log \frac{4 \alpha^2}{\Lambda^2} \right ) + \mbox{s.t.} \,,
\label{eq:II4}
\ee
with 
\be
A=4 \alpha^2 + (1-x) \mu^2 + P^2 x(1-x) \,,
\label{eq:A}
\ee
and the limit $\mu^2 \rightarrow 0$ can be taken without any ambiguity. 
Note that in $I_2$ we have kept the momentum cut-off $\Lambda$ to stress that $I_2$ is the term that leads to an IR
singularity at zero external momentum and hence to the well known effect of UV/IR mixing. There is no UV/IR mixing from $I_3$ and
$I_4$ as these integrals vanish at zero external momentum. This is rather interesting, because $I_3$ and $I_4$ appear only due to the noncommutativity of the theory (they vanish identically at $\theta = 0$), nevertheless they do not lead to UV/IR mixing. However, they diverge logarithmically. Moreover, it is worthwhile to note that in (\ref{eq:II3}) and (\ref{eq:II4}) (or more precisely in (\ref{eq:III3}) and (\ref{eq:III4})), the $\theta \rightarrow 0$ limit should be taken along with $\Lambda \rightarrow \infty$ to obtain the correct result. This is so because, when $\theta \neq 0$, it is necessary to regularize the integrals in (\ref{eq:I3}) and (\ref{eq:I4}) by suitably introducing the cut-off $\Lambda$. As $\theta \rightarrow 0$ this cut-off is no longer required, and it must be removed as the integrands in (\ref{eq:I3}) and (\ref{eq:I4}) vanish identically.

For the field $\rho$, the one-loop two-point function can now be expressed as
\be
I_\rho(P^2) := \langle \rho(P) \rho(P) \rangle : =  \left (\frac{1}{2} I_3  + I_4 \right) \Big|_{4 \alpha^2 \rightarrow \mu^2} \,. 
\ee
Thus we observe that $\langle \rho(P) \rho(P) \rangle$ is present purely due to the noncommutativity of the theory, but amusingly it does not lead to any UV/IR mixing. 

Let us now briefly discuss the mass and the field strength 
renormalization in the Euclidean signature. The renormalized self-energy of $\varphi$ can be given as 
\be
\Sigma_{R}(P^2) = Z_\varphi^{-1} I(P^2) + \delta m^2_\varphi - \delta Z_\varphi P^2 \,,
\ee
where $Z_\varphi = 1 + \delta Z_\varphi$. We can determine $\delta m^2_\varphi$ and $\delta Z_\varphi$
from the renormalization conditions
\be 
\Sigma_R(P^2) \Big |_{P^2 =P_0^2} = 0 \,, \quad \frac{d}{d P^2}
\Sigma_R(P^2) \Big |_{P^2 =P_0^2} = 0 \,
\ee   
at an arbitrary reference momentum $P_0^2$. 

For instance, when $\Lambda \rightarrow \infty$, we can focus on the logarithmically divergent parts of $I_\varphi(P^2)$ and $I_\rho(P^2)$. For $\delta m^2_\varphi$ and $\delta Z_\varphi$ these conditions lead to
\be
\delta m^2_\varphi = \frac{1}{1 + \delta Z_\varphi} \left [ \frac{\alpha^2}{3 \pi} \log \frac{4 \alpha^2}{\Lambda^2} \right ] \,, \quad
\delta Z_\varphi = \frac{-1 + \sqrt{1- \frac{7}{2^4 \pi} \log \frac{4 \alpha^2}{\Lambda^2}} }{2} \,.
\ee

A similar calculation shows that there is no mass renormalization for the field $\rho$, and the field-strength renormalization is given by
\be
\delta Z_\rho = \frac{-1 + \sqrt{1- \frac{5}{2^5 \pi} \log \frac{\mu^2}{\Lambda^2}} }{2} \,.
\ee
It is important to stress, that the above expressions for $\delta m^2_\varphi$, $\delta Z_\varphi$ and $\delta Z_\rho$ are valid for $\theta \neq 0$, although $\theta$ does not explicitly appear in them. As we have already remarked, when $\theta$ approaches to zero in $I_\varphi(P^2)$ and $I_\rho(P^2)$, the divergent terms in the cut-off $\Lambda$ cancel with those terms divergent in $\theta$. In this case, the standard answer for the commutative sine-Gordon model will be recovered, and a mass counter term will be sufficent to renormalize the theory.

When the results for the one-loop amplitude $I(P^2)$ are analytically continued to the
Minkowski space, the logarithms develop branch cuts. This leads to imaginary parts in the total 
one-loop amplitudes $I_\varphi(P^2), I_\rho(P^2)$, which for space-like external momenta leads to a violation of unitarity. 
The latter is a rather typical behaviour, known to occur in certain formulations of time-space noncommutative field theories \cite{Gomis}. We observe the integrability and causal tree-level S-matrix of the current model are unable to improve this rather catastrophic behaviour.

It is, however, important to point out that there are alternative ways of formulating time-space noncommutative theories, which preserve unitarity \cite{Doplicher} \cite{Balachandran1} \cite{Balachandran2}. The applicability of these formulations to the present model remains an open problem. In this context, we note that the integrability of a time-space noncommutative sinh-Gordon model has recently been studied in \cite{Vaidya}.

\section{Conclusions and Outlook}

In this article we employed semi-classical methods to study the quantum properties of the integrable time-space noncommutative sine-Gordon model defined by the action (\ref{eq:NCSGaction}). We have examined the fluctuations at quadratic order around the static kink solution. The spectrum of the fluctuations for the sine-Gordon field consists of a single discrete mode (the zero mode) followed by a continuum. Applying standard perturbation theory, we have proved that at $O(\theta)$ this spectrum coincides with that of the corresponding commutative theory. We have also reasoned, by means of qualitative arguments, that the same conclusion holds at $O (\theta^2)$ as well. It is worthwhile to note that the collective-coordinate quantization of the zero mode may reveal novel properties of this model. However, this appears to be a formidable task, as the standard methods are not directly applicable in this context, due to time-space noncommutativity. 

We also studied the one-loop structure of the two-point functions for the sine-Gordon field $\varphi$ and the additional scalar  field $\rho$, in the vacuum sector and showed that they have logarithmic divergences. Using these results, we have computed the mass and field strength renormalization counterterms in the Euclidean signature. We have seen that the two-point function for the sine-Gordon field exhibits UV/IR mixing, and one-loop amplitudes for both $\varphi$ and $\rho$ develop imaginary parts under Wick rotation to the Minkowski signature. The latter fact violates unitarity relations for space-like external momenta. This property presents an important obstacle in studying the quantum corrections to the mass of the sine-Gordon kink. Although the usual vacuum subtraction can be performed as in (\ref{eq:vacsubtract}), the mass and field-strength renormalization counterterms can not be unambiguously identified in Minkowski space. It may be useful to study the corresponding aspects of the $2 + 1$-dimensional sigma model \cite{Lechtenfeld:2001gf} from which the model considered in this paper descends by dimensional and algebraic reduction. This may help us to built further inroads to the structure of these theories. \\[0.3em]
{\bf Acknowledgements}\\[0.2em]
S.K. is supported by the Deutsche Forschungsgemeinschaft (DFG) under grant LE 838/9.

\appendices

\subsection{Feynman rules}

In Euclidean signature, Feynman rules that follow from the action (\ref{eq:quadraticaction}) are as follows.
In all the vertices momentum conservation is already imposed.
\begin{itemize}
\item{The propagators are
\be
\parbox{2cm}{\includegraphics[width=1.5cm]{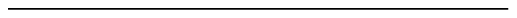}}
\equiv \langle \varphi \varphi \rangle = \frac{2 }{k^2 + 
 4\alpha^2} \,, \quad \quad
\parbox{2cm}{\includegraphics[width=1.9cm]{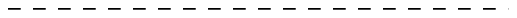}}
\equiv\ \langle \rho\,\rho \rangle = \frac{2}{k^2} \,.
\ee}
\item{The vertices are
\beqa
\parbox{2cm}{\includegraphics[width=1.5cm]{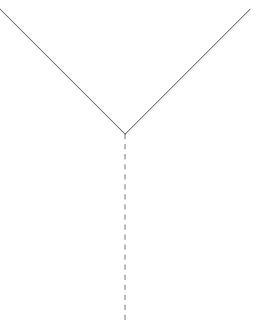}}
&=& -\frac{1}{2^2} \Big [ (k_1 \wedge k_2) \sin \big (\theta \frac{k_1 \wedge k_2}{2} \big )  \\[-1em]   
&& \quad \hskip 0.2cm + (k_2 \wedge k_3) \sin \big (\theta \frac{k_2 \wedge k_3}{2} \big ) + (k_1 \wedge k_3) \sin \big (\theta \frac{k_1 \wedge k_3}{2} \big ) \Big ] \, \nn \\
\parbox{2cm}{\includegraphics[width=1.5cm]{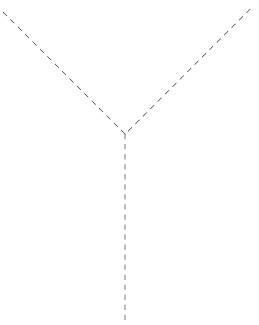}}
&=& -\frac{1}{3 \cdot 2^2} \Big [ (k_1 \wedge k_2) \sin \big (\theta \frac{k_1 \wedge k_2}{2} \big )  \\[-1em]   
&& \quad \quad \hskip 0.3cm + (k_2 \wedge k_3) \sin \big (\theta \frac{k_2 \wedge k_3}{2} \big ) + (k_1 \wedge k_3) \sin \big (\theta \frac{k_1 \wedge k_3}{2} \big ) \Big ] \, \nn
\eeqa} 
\beqa
\parbox{2cm}{\includegraphics[width=1.5cm]{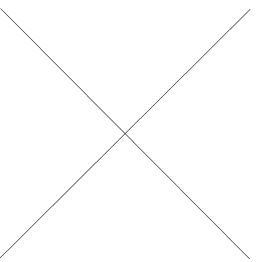}}
&=& - \frac{i}{2^4\cdot4!} \Big [ k_1 \cdot (k_3 - k_2) \sin \big(\theta \frac{ k_2 \wedge k_3}{2}\big) 
e^{-\frac{i}{2} \theta k_1 \wedge (k_2 + k_3 )} \\[-1em]
&& \quad \quad \quad + \, k_2 \cdot (k_4 - k_3) \sin \big(\theta \frac{ k_3 \wedge k_4}{2}\big) e^{-\frac{i}{2} \theta k_2 \wedge (k_3 + k_4 )} \nn \\
&& \quad \quad \quad  + \, k_3 \cdot (k_1 - k_4) \sin \big(\theta \frac{ k_4 \wedge k_1}{2}\big) e^{-\frac{i}{2} \theta k_3 \wedge (k_1 + k_4 )}  \nn \\
&& \quad \quad \quad  + \, k_4 \cdot (k_2 - k_1) \sin \big(\theta \frac{ k_1 \wedge k_2}{2}\big) e^{-\frac{i}{2} \theta k_4 \wedge (k_1 + k_2 )}
\Big ] \nn \\
&&\quad \quad \quad \quad \hskip 1.9cm + \frac{1}{12} \alpha^2 e^{ -\frac{i}{2} \theta (k_1 \wedge k_2 + k_1 \wedge k_3 + k_2 \wedge k_3) } \, \nn \\
\parbox{2cm}{\includegraphics[width=1.5cm]{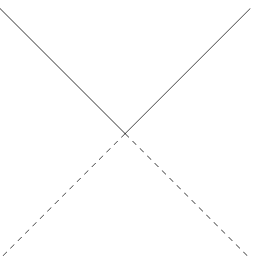}}
& =& \frac{1}{2^3 4!}  \Big [ (k_1 -k_2) \cdot (k_3 - k_4)
\sin\big(\theta \frac{ k_1 \wedge k_2}{2}\big) \, \sin \big
  (\theta \frac{k_3 \wedge k_4}{2} \big) \\[-2pt]
&& \hskip 0.9cm (k_2 -k_3) \cdot (k_4 - k_1)
\sin\big(\theta \frac{ k_2 \wedge k_3}{2}\big) \, \sin \big
  (\theta \frac{k_4 \wedge k_1}{2} \big) \Big ] \nn \\
\parbox{2cm}{\includegraphics[width=1.5cm]{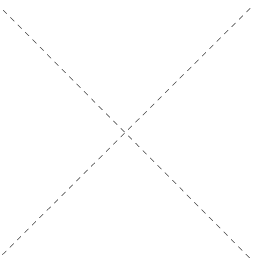}}
&=& - \frac{i}{2^4\cdot4!} \Big [ k_1 \cdot (k_3 - k_2) \sin \big(\theta \frac{ k_2 \wedge k_3}{2}\big) 
e^{-\frac{i}{2} \theta k_1 \wedge (k_2 + k_3 )} \\[-1em]
&& \quad \quad \quad + \, k_2 \cdot (k_4 - k_3) \sin \big(\theta \frac{ k_3 \wedge k_4}{2}\big) e^{-\frac{i}{2} \theta k_2 \wedge (k_3 + k_4 )} \nn \\
&& \quad \quad  \quad + \, k_3 \cdot (k_1 - k_4) \sin \big(\theta \frac{ k_4 \wedge k_1}{2}\big) e^{-\frac{i}{2} \theta k_3 \wedge (k_1 + k_4 )}  \nn \\
&& \quad \quad \quad + \, k_4 \cdot (k_2 - k_1) \sin \big(\theta \frac{ k_1 \wedge k_2}{2}\big) e^{-\frac{i}{2} \theta k_4 \wedge (k_1 + k_2 )}
\Big ] \nn \,.
\eeqa 
\end{itemize}

\subsection{Results of the loop integrals}

In this Appendix we give the full result for the integrals $I_1, I_2,
I_3, I_4$ given in (\ref{eq:I12}), (\ref{eq:I3}), (\ref{eq:I4}). We have
\be
I_1 = \frac{2 \alpha^2}{3 \pi} K_0 \left ( \frac{4 \alpha}{\Lambda}
\right ) \,, \quad 
I_2 = \frac{\alpha^2}{3 \pi} K_0 \left ( 4 \alpha \sqrt{\frac{\theta^2 P^2}{4} + \frac{1}{\Lambda^2}} \right ) \,,
\ee
\be
I_3 = \frac{3^2}{2^4 \pi} \int_0^1 d x \Big [ -\frac{P^2}{2} K_0 (
\theta P \sqrt{A}) - 2 P \sqrt{A} K_1 (\theta P \sqrt{A}) 
+ \frac{P^2}{2} K_0 \left( \frac{2 \sqrt{A}}{\Lambda} \right) \Big ] \,,
\label{eq:III3}
\ee
where $P = \sqrt{P^2}$, $K_\nu(x)$ is the modified Bessel function and $A$ is
already defined in (\ref{eq:A}).

For $I_4$ we have
\be
I_4 = \frac{1}{2^4 \pi} \Bigg [- P^2 K_0 \left (\frac{4 \alpha}{\Lambda}
\right ) + P^2 K_0 ( 2 \alpha \theta P) - 4 \alpha^2 K_{-2}( 2 \alpha \theta P) 
+ \frac{2 \alpha}{\theta P} K_{-1}(2 \alpha \theta P ) \Bigg ] \,.
\label{eq:III4}
\ee
It is worthwhile to note that an integral of the form 
$\frac{1}{(2 \pi)^2} \int d^2 k \frac{k^2}{k^2 + 4 \alpha^2}$ is present in
(\ref{eq:I4}). This integral can be set to zero after dimensional
regularization. In order to see this, note that in $d$ dimensions we have
\be
\frac{1}{(2 \pi)^2} \int d^d k \frac{k^2}{k^2 + 4 \alpha^2} = \frac{- i
  \frac{d}{2} \Gamma(- \frac{d}{2})}{(4 \pi)^{\frac{d}{2}} (4
  \alpha^2)^{- \frac{d}{2}}} \,.
\ee
This expression has no poles at $d=0$, and for $d \geq 0$ it is
proportional to a positive power of $4 \alpha^2$. Thus, it can be set 
to zero without loss of generality.

\end{document}